\documentclass[aps,prl,preprint,floatfix,superscriptaddress]{revtex4-1}
\usepackage{amsmath,amssymb}
\usepackage{graphicx}
\usepackage[utf8]{inputenc}
\usepackage[T1]{fontenc}
\usepackage[english]{babel} 
\usepackage{color}
\usepackage{units}
\usepackage[group-separator={,}]{siunitx}
\usepackage{hyperref}
\hypersetup{colorlinks=true,citecolor={blue},linkcolor={blue},urlcolor={blue}}
\usepackage[note-name=,use-sort-key=false]{notes2bib}
\usepackage{xr}
\usepackage{xcolor}
\usepackage{setspace}

\usepackage{fullpage}

\begin{document}

\title{Dissipative phase locking of exciton-polariton condensates}

\author{H. Ohadi}
\email[correspondence address: ]{hamid.ohadi@soton.ac.uk}
\author{R. L. Gregory}
\author{T. Freegarde}
\affiliation{
 School of Physics and Astronomy, University of Southampton, Southampton, SO17 1BJ, United Kingdom
}
\author{Y. G. Rubo}
\affiliation{
Instituto de Energías Renovables, Universidad Nacional Autónoma de México, Temixco, Morelos, 62580, Mexico
}
\author{A. V. Kavokin}
\affiliation{
 School of Physics and Astronomy, University of Southampton, Southampton, SO17 1BJ, United Kingdom
}
\affiliation{
Spin Optics Laboratory, St. Petersburg State University, 1, Ulianovskaya, St. Petersburg, 198504, Russia
}
\author{P. G. Lagoudakis}
\affiliation{
 School of Physics and Astronomy, University of Southampton, Southampton, SO17 1BJ, United Kingdom
}

\begin{abstract}
We demonstrate, both experimentally and theoretically, a new phenomenon: the presence of dissipative coupling in the system of driven bosons. This is evidenced for a particular case of externally excited spots of exciton-polariton condensates in semiconductor microcavities. We observe that for two spatially separated condensates the dissipative coupling leads to the phase locking, either in-phase or out-of-phase, between the condensates. The effect depends on the distance between the condensates. For several excited spots, we observe the appearance of spontaneous vorticity in the system.
\end{abstract}

\maketitle

Exciton-polaritons (polaritons) are bosonic quasiparticles formed by the strong
coupling of photons in a Fabry-Perot microcavity with excitons in a
semiconductor quantum well~\cite{kavokin_microcavities_2007}.  Due to their
finite lifetime, polaritons need to be externally pumped.  Once the income rate
of polaritons exceeds their decay rate (i.e. the threshold), a condensate with
macroscopic occupation is
formed~\cite{kasprzak_bose-einstein_2006,balili_bose-einstein_2007,
    deng_condensation_2002,schneider_electrically_2013}. Condensation of
exciton-polaritons is an example of a spontaneous symmetry breaking process in a
many-body
system~\cite{snoke_spontaneous_2002,baumberg_spontaneous_2008,ohadi_spontaneous_2012}.
Being externally driven, polaritons can condense in one or several quantum
states out of thermal equilibrium~\cite{kasprzak_formation_2008}, in contrast to
atomic Bose-Einstein
condensates~\cite{davis_bose-einstein_1995,anderson_observation_1995}. Once
polaritons are condensed they flow out of the pump spots and experience an
increase in in-plane momentum due to their repulsion from hot excitons in the
reservoir and other
polaritons~\cite{richard_spontaneous_2005,wouters_spatial_2008,wertz_spontaneous_2010}.
The outflowing polaritons may couple and phase lock the spatially separated
condensates~\cite{tosi_geometrically_2012}. The coupling mechanism has been
attributed to the coherent ``ballistic coupling'' mechanism, whereby each
condensate center is resonantly pumped by the outflow from the neighboring
condensates.

Here, we investigate in detail the coupling mechanism of spatially separated
condensates using a simple two-condensate geometry and find that the coherent
``ballistic coupling'' picture is inadequate in describing the phase locking of
spatially separated polariton condensates.  We show how two condensates could
phase lock in symmetric or antisymmetric states depending on their separation
distance, as well as their outflowing condensate wavevectors. Using an
incoherent coupling mechanism, to which we refer as \textit{dissipative
    coupling} here, we explain this behavior and generalize it to any array
geometry in particular a triangular condensate array. We show experimentally and
theoretically how dissipative coupling could result in macroscopic pure states
where all condensates are in phase or nontrivial mixed states where adjacent
condensates have $\pm2\pi/3$ phase difference giving rise to the spontaneous
appearance of vortices at the center of the array, where the condensates
overlap.

\begin{figure}
\centering
\includegraphics[width=0.56\textwidth]{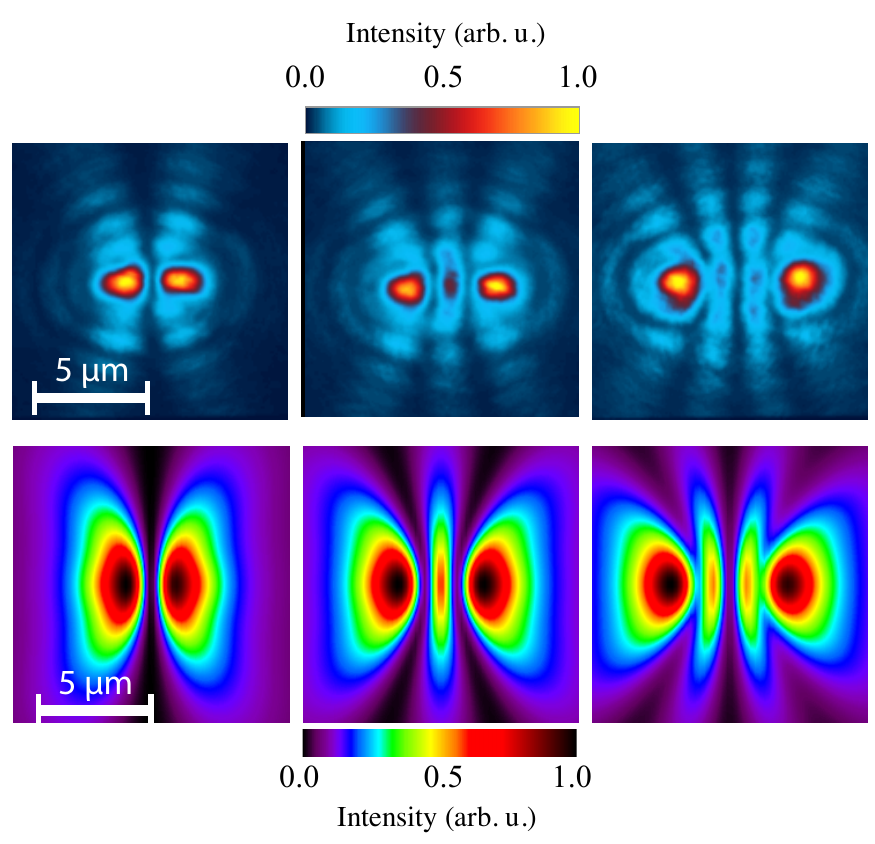}
\caption{(upper row) The
time-integrated images of two polariton condensates in real space at various
separation. The condensates were created simultaneously at $P\simeq P_{th}$,
where $P_{th}$ is the threshold power for condensation.  The observation of an
interference pattern between the two condensates indicates that they are
phase correlated.  Zero intensity of the interference pattern in the middle of
the two sources indicates that the two condensates are anti-synchronized. (lower
row) The time-integrated GP simulations with random initial phase for 75
realizations at separation distances between the condensates as in the experiment is shown.
The condensates phase lock on average to zero or $\pi$ phase difference
depending on their separation distance.}
\label{fig:cavity-phase}
\end{figure}

\begin{figure}
\centering
\includegraphics[width=0.56\textwidth]{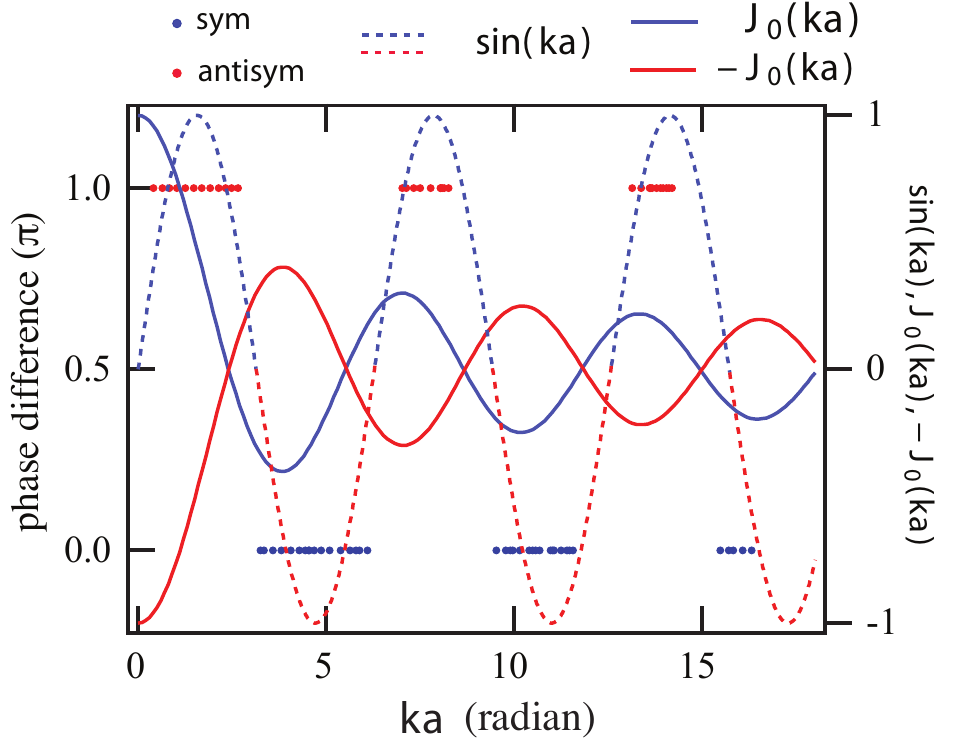}
\caption{
The condensates flip from
symmetric to antisymmetric state as their separation ($a$) and outflow polariton
in-plane wavevector ($k_c$) changes. The red circles show the antisymmetric and
the blue circles show the symmetric states (see also Video S1). The
solid curves are the calculated losses due to emission of free polaritons for
symmetric (blue) and antisymmetric (red) states versus the unitless parameter
$k_ca$. The dashed line which is plotting $\sin(k_ca)$ shows how the phase
difference will behave if the coupling is coherent; the blue dashed lines are
for symmetric phase locking and red for antisymmetric.}
\label{fig:bessel_curves}
\end{figure}

\begin{figure*}
\centering
\includegraphics[width=1\textwidth]{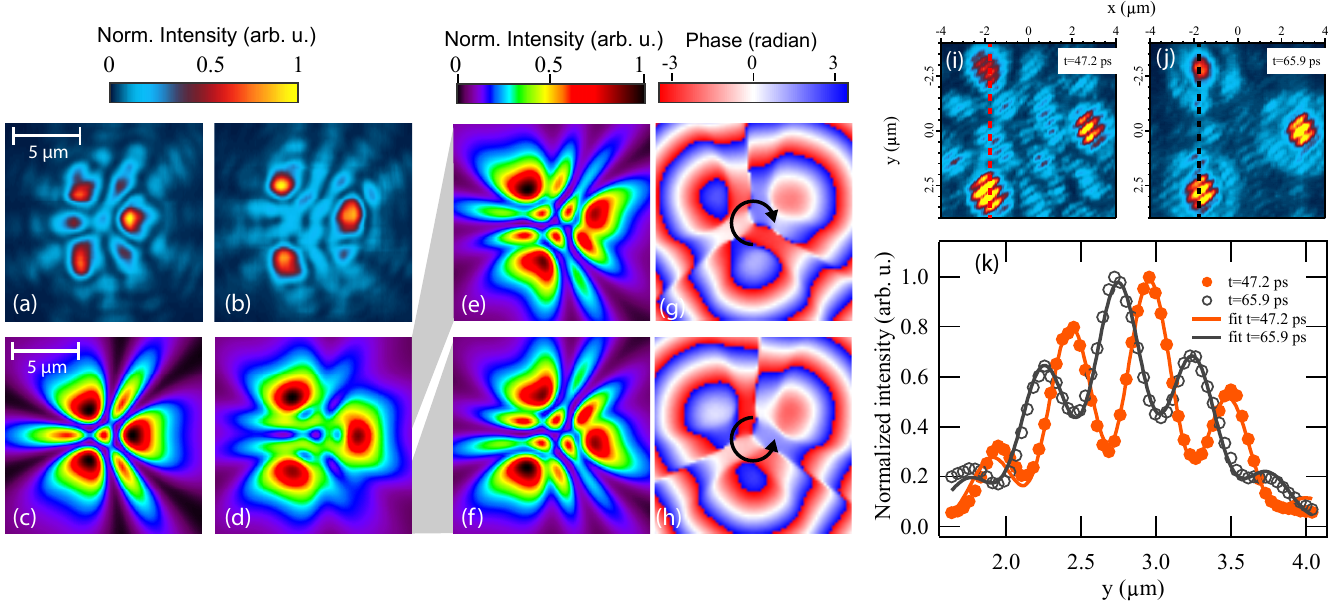}
\caption{(a-d) Time-integrated real-space tomography images
of the triangular array condensates are shown for $a=\unit[4]{\mu m}$ (a),
$a=\unit[5.5]{\mu m}$ (b). (c,d) The GP simulations with
random initial phase, averaged over 75 realizations at different separations matching
the experimental conditions are shown. In (c) all three condensates are in phase
but (d) is a mixture of vortices with $\pm1$ winding numbers shown in (e)
and (f). The phase diagrams of (e) and (f) are shown in (g) and (h) proving that
they are indeed vortex states with winding numbers of $+1$ (h) or $-1$ (g).
(i,j) The time resolved interferograms resulted from the
interference of two condensates with each other are shown at $t=\unit[47.2]{ps}$
(i) and $t=\unit[65.9]{ps}$ (j).  (k) The circles show the line
profiles of (i) and (j) taken at $x=\unit[-1.9]{\mu m}$ (shown by the dashed
lines). The solid lines are the
fits by the convolution of a Gaussian function with $\cos(ky+\phi)$ resulting in
a $(0.78 \pm 0.02) \times \pi$ phase difference between the two fits. }
\label{fig:3beams}
\end{figure*}

We create polariton condensates by nonresonant pulsed excitation in a
semiconductor microcavity~(see Supp. Info.~\ref{sec:methods}). By studying the
near-field photoluminescence (PL) spectra of the cavity we can measure the phase
and the momentum of the polaritons in each condensate. A condensate array is
created with each excitation pulse by means of optical shaping of the pump beam.
To gain more insight into the phase locking mechanism, we study the simplest
case of a pair of condensates. Fig.~\ref{fig:cavity-phase} shows the
time-integrated real space PL of a condensate pair at different spatial
separations. The pattern resembles the interference pattern of two phase-locked
cylindrical waves with wavevectors equal to the in-plane wavevector of free
polariton eigenmodes resonant in frequency with the condensates~\footnote{This
    pattern may look similar to the work reported in
    Ref.~\citenum{tosi_sculpting_2012} but the physics behind the observed
    patterns is different in these two works. In our case it is because of the
    interference of two phase-locked sources rather than the confinement of the
    wave function in a parabolic potential. This is very clear as our energy
    resolved real-space images show that the fringes are at the same energy as
    the condensates are~(see Supp.  Fig.~\ref{fig:rs-er}).  Also all real-space
    images shown in Fig.~\ref{fig:3beams} are tomography images taken at the
    same energy as the condensates.}. The appearance of an interference pattern
indicates that the condensate pairs are phase-locked at zero or $\pi$ phase
difference depending on their separation or outflow in-plane wavevector. The
phase difference has a nearly periodic dependence on the product of the
condensate pair separation and the wavevector of outflowing polaritons ($k_c$),
as shown in Fig.~\ref{fig:bessel_curves} by blue and red dots (see also Supp.
Fig.~\ref{fig:ks-rs} and Video S1).  Their relative phase changes abruptly as a
function of the distance, which is untypical for the conventional Josephson
coupling~\cite{lagoudakis_coherent_2010}. The coherent coupling mechanism
requires in-phase coupling when the separation between two condensates is
smaller than half the in-plane wavelength of the condensate ($\pi/k_c$), as
shown by the dashed lines in Fig.~\ref{fig:bessel_curves}. This is opposite to
what we observe and also opposite to what our Gross-Pitaevskii simulations
reveal (see Supp.  Fig.~\ref{fig:phase-diagram}).

The phase-locking mechanism can be understood if we consider condensation as an
inherently dissipative and symmetry breaking process. Since the excitation is
nonresonant, the phase coherence from the laser is lost and polaritons are
initially created with random phases uniformly distributed across the pump spot.
Due to phase fluctuations at the onset of condensation, this phase symmetry
breaks down, and a macroscopic phase is built up along the whole condensate. One
must consider the evolution of the unified wave function for both condensates,
which naturally includes their interaction as they expand and interfere with
each other.  Condensed polaritons are repelled and ejected from the condensates
with a specific in-plane wavevector $k_c$ because of the interactions with each
other and more importantly with the excitons in the
reservoir~\cite{askitopoulos_polariton_2013} (see also Supp.
Fig.~\ref{fig:blueshift-schem}). There are two types of losses from the
condensates centers: (i) cavity losses through the Bragg mirrors counted by
polaritons' radiative decay rate which could result in the radiative coupling of
individual condensates~\cite{aleiner_radiative_2012}, and (ii) losses due to the
in-plane flow of polaritons with the wavevector $k_c$ away from the excitonic
reservoir, which is discussed here.  Outflowing polaritons from different
condensates can interfere constructively or destructively depending on the
phases that they gain during their flow. The configurations which provide a
destructive interference between the outflowing polaritons result in lower
losses of polaritons from the condensates and higher polariton occupation
numbers. This is amplified by stimulated scattering of polaritons from the
reservoirs, which eventually leads to phase locking of condensates. In other
words, interference between the two condensates breaks their individual phase
symmetries to a state with the maximum wave function occupation number.  Because
the interference depends on the phase gained during the flow from one condensate
to another, the phase relation between these phase-locked condensates depends on
their positions or more generally on their topology as well as on the in-plane
wavevector of outflowing polaritons. This coupling mechanism is similar to the
phase locking of Huygens's
clocks~\cite{huygens_oeuvres_1888,bennett_huygenss_2002}. If the two clocks
pendulums swing in in-phase mode, they tend to push the frame in the same
direction resulting in frictional forces that eventually dampen the motion of
the pendulums. If they swing in anti-phase mode the back actions cancel out and
the frame does not move, minimizing the dissipative losses. The same mechanism
is responsible for the sustained aftersound of the
pianoforte~\cite{weinreich_coupled_1977}. The dissipative mechanism due to
interference also closely resembles two radiating dipoles, where the loss (power
dissipation) depends on their relative phase and
separation~\cite{smith_insightful_2010}. In our system dissipation is governed
by the interference of bosons outflowing from the condensates' centers.

The dependence of the total losses of two condensates on the phase difference $\theta$
between the individual condensates is given by
\begin{equation}
 \label{eq:tau_of_theta}
 I(k_ca,\theta) \propto [1 +J_{0}(k_{c}a) \cos(\theta)],
\end{equation}
where $J_0$ is the Bessel function (see Supp. Info.~\ref{si:coupling}).  This
simple relation defines the dependence of the total losses of the polariton
condensates on the phase difference between the individual condensates. At the
formation of the condensates, the phase symmetry is spontaneously broken to a
state that maximizes the total occupation number, which is a state that
minimizes the losses. The blue and red solid lines in
Fig.~\ref{fig:bessel_curves} shows the calculated losses of the condensates
($I(k_ca,\theta)$) for symmetric and antisymmetric states versus separation
between the two condensates respectively. At each separation, the condensates
pick on average the state with lower losses and flip between in-phase and
anti-phase states in a nearly periodic fashion as their separation changes.  The
lower panel in Fig.~\ref{fig:cavity-phase} shows the 2D GP simulations with
random initial conditions which is coupled to a hot exciton reservoir excited by the
nonresonant pump~\cite{wouters_spatial_2008}. Each figure is a time-integrated
average of 75 realizations. The fringes between
the condensates are due to interference between condensates that are phase
locked on average in either zero or $\pi$ phase difference. The phase difference
depends on their separation and in-plane wavevector as the result of the phase
symmetry breaking at the onset of condensation. 

\begin{figure}[h]
\centering
\includegraphics[width=0.55\textwidth]{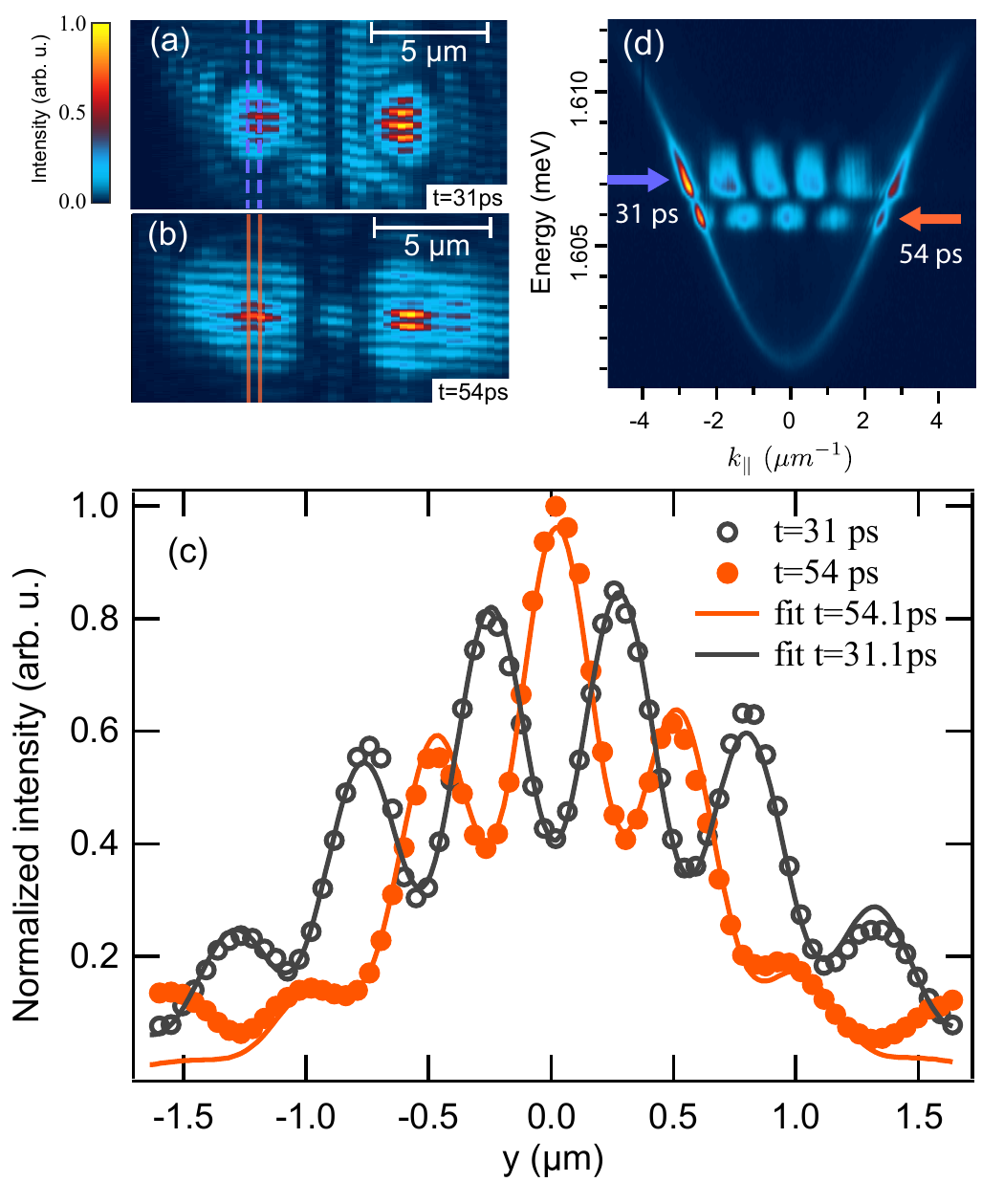} \caption{(a,b) The time-resolved interferograms at different times
showing the phase synchronization of the two condensates. At early times and
before condensation, they are not phase locked. When the condensates are formed
they anti-synchronize (a) and as they relax in energy they synchronize (b).
(c) The intensity profile of the interferograms at the position marked
by the vertical lines in (a) and (b) at different times. The phase difference
between the two condensates changes by $\pi$ as they relax down in energy. The
solid lines are the fits of a Gaussian function convoluted with a $\cos(k
y+\phi)$, giving a phase difference of $(0.98 \pm 0.01) \times \pi$ between the two states.
(d) The time-integrated momentum space is showing that condensation occurs at
higher energies, with a gradual energy relaxation as time passes. At $t=\unit[31]{ps}$
(marked by a blue arrow) the two condensates are anti-phase synchronized and
they synchronize in phase at $t=\unit[54]{ps}$ (marked by an orange arrow).} 
\label{fig:TR}
\end{figure}

We observe a  transition from the in-phase state to the out-of-phase state in
time. By increasing the pump power to nearly twice the threshold power, we form
condensates at a higher energy and larger in-plane wavevectors. We perform
time-resolved interferometry of the two condensates by interfering the PL from
one condensate with that from the other one using an actively stabilized
Michelson interferometer in a mirror-retroreflector configuration (for details
see Ref.~\citenum{kasprzak_bose-einstein_2006} and Supp.  Fig.~\ref{fig:setup}).
From the interferograms we can extract the time-resolved relative phase between
them (see also Video S2).  Fig.~\ref{fig:TR}(a) shows the interference fringes
just after the condensation.  Initially, the two condensates anti-synchronize
($\pi$ out of phase). As they relax in energy by phonon and exciton scatterings,
they desynchronize and resynchronize at a later time [Fig.~\ref{fig:TR}(b)].
However, when the two condensates resynchronize their phase difference changes
by $\pi$ [blue and orange line in Fig.~\ref{fig:TR}(c)]. The symmetry flipping
can also be seen in the momentum space images shown in Fig.~\ref{fig:TR}(d).
The condensates first anti-synchronize at a high energy marked by the blue arrow
($t=\unit[31]{ps}$) and then synchronize at a later time ($t=\unit[54]{ps}$)
marked by the orange arrow.

Fig.~\ref{fig:3beams}(a,b) show the spectral tomography images of an
equilateral triangular condensate arrays at two different array parameters
(separation). Observation of dark and bright fringes at the same energy as the
condensates themselves indicates that the condensates are phase locked.
Constructive interference at the center of the array corresponds to the case
where they are all in phase, and destructive interference corresponds to the
case where they are out of phase.  Fig.~\ref{fig:3beams}(c,d) are the
GP simulations with random initial phase (see Supp. Info.
\ref{sec:GP}) for 75 realizations, which demonstrate the in-phase (c) and
out-of-phase locking of the condensates (d). Here the losses are minimized if
either all three neighbors are in phase or if there is a $2 \pi/3$ phase
difference between the neighbors (see Supp. Info.~\ref{si:coupling} and Supp.
Fig.~\ref{fig:3beams-geometry}). The nontrivial case where there is a $2\pi/3$
phase difference going clockwise or anticlockwise between the neighbors is shown
in Fig~\ref{fig:3beams}(d). Fig.~\ref{fig:3beams}(d) is composed of two
topologically different nondegenerate states (shown in panel E and F), with
clockwise and anticlockwise $2\pi/3$ phase difference between adjacent
neighbors. They correspond to phase vortices with winding numbers of $+1$ or
$-1$ (shown in G and H). Zero intensity in the center of the system is present
for both types of vortices. The condensates switch between all-in-phase to
vortex states in a nearly periodic manner as the array parameter changes (see
Supp.  Fig.~\ref{fig:3beams-larger-sep} for larger separations). 

A dynamical transition in time from an all-in-phase state to out-of-phase state
can be observed if we keep the array parameter constant and pump the microcavity
at higher powers (see Video S3). We pump the cavity at nearly twice the
threshold power for condensation. We then interfere the PL from one condensate
with that from the other one using the Michelson
interferometer.
Fig.~\ref{fig:3beams}(i,j) show the interferograms at different times when
the condensates are all in-phase (i) and when they are out-of-phase (j) (see
also Video S4).  The line profiles taken from these two interferograms, shown in
Fig.~\ref{fig:3beams}(k), show that these two states are out of phase by nearly
$2\pi/3$ (see also Fig.~\ref{fig:3cond-kspace} for the momentum space). The
interferometry measurements show that any two neighboring condensates are phase
locked. There are two possible stable states in the system: either all
condensates are in phase or there is a $\pm2\pi/3$ phase difference between
neighbors, which correspond to clockwise or anticlockwise vortices with $\pm1$
winding numbers. This dynamical transition from an all in-phase state to a
vortex state could also be observed in the momentum space.  The transition is
caused by the change of condensates outflowing polaritons wavevector as the
condensates relax in energy by phonon and exciton
scatterings~\cite{wouters_energy_2010}.

In conclusion,  we studied the phase coupling mechanism of spatially separated
polariton condensate pairs. We demonstrated theoretically and experimentally
that depending on the wave vector of the outflowing polaritons and the
separation of the condensates, a pair of condensates phase synchronize with zero
or $\pi$ phase difference. We explained how losses due to the outflowing
polaritons from the condensates are minimized for states with a specific phase
difference.  Our simulations showed how the spontaneous formation of these
vortices is directly related to the spontaneous symmetry breaking and the
nonlinearity at the condensation phase transition.  We demonstrated the
\textit{spontaneous} vorticity resulting from the phase locking of a triangular
condensate array.  The condensate array spontaneously picks a clockwise or
anticlockwise rotating direction due to phase fluctuations at the onset of
condensation. Our observations demonstrate the spontaneous symmetry breaking in
space in a dissipative bosonic system.

\begin{acknowledgments}
The authors thank Jacqueline Bloch and Aristide Lema\^{\i}tre for the
  provision of the sample. H.O. and A.V.K would like to thank Jacqueline Bloch
  for stimulating discussions.  P.G.L. and A.V.K. would like to acknowledge the
  Royal Society and EPSRC through contract EP/F026455/1 for funding. AK
  acknowledges support from Russian Ministry of Education and Science (contract
  No.11.G34.31.0067).  H.O.  acknowledges the use of the IRIDIS High Performance
  Computing Facility, at the University of Southampton.
\end{acknowledgments}

\clearpage
\setcounter{figure}{0}
\makeatletter 
\renewcommand{\thefigure}{S\@arabic\c@figure}
\makeatother

\setcounter{secnumdepth}{2} 
\setcounter{section}{0}

\section{Supplemental Information}
\subsection{Experimental setup}
\label{sec:methods}

\begin{figure}
\centering
\includegraphics[width=0.6\textwidth]{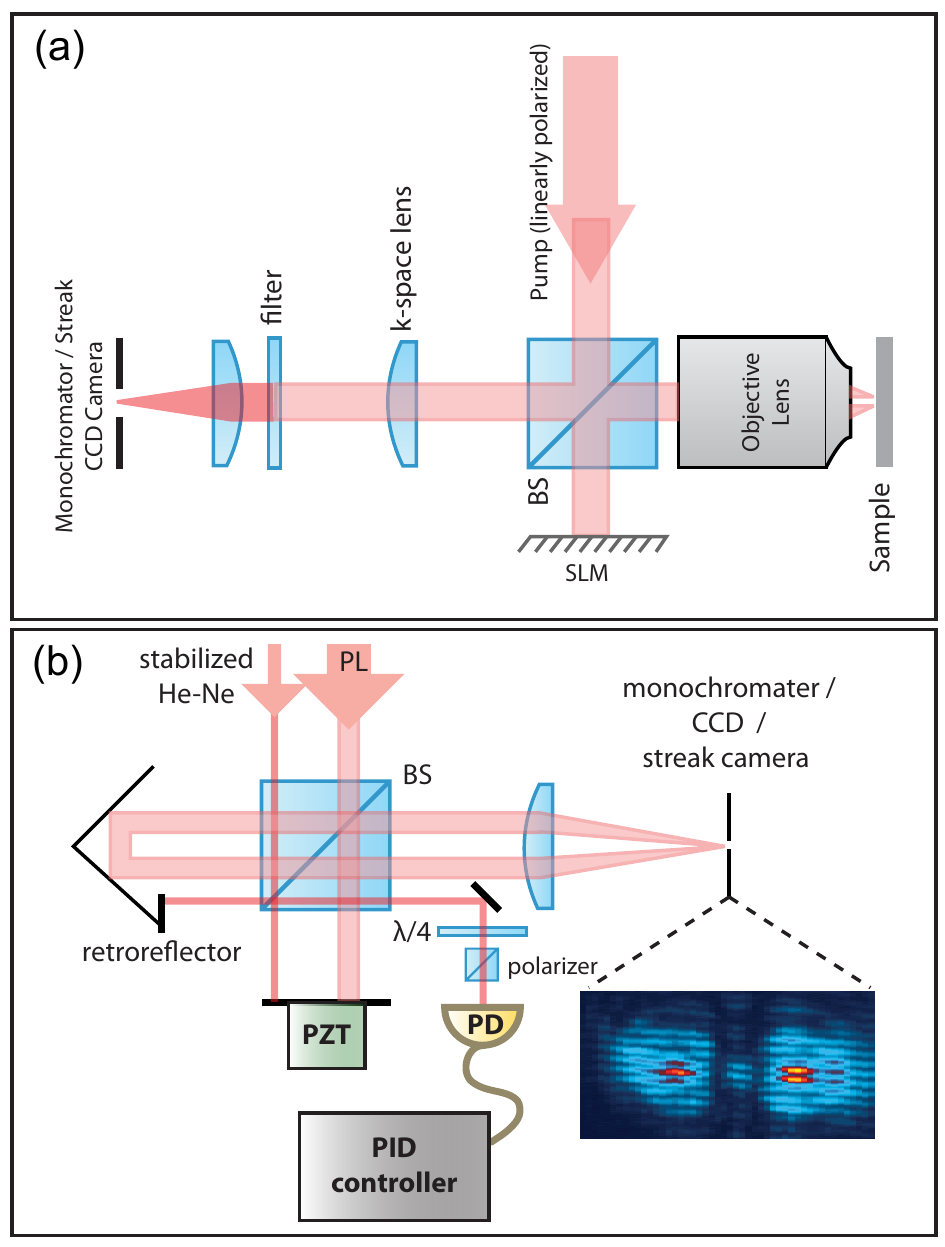}
\caption{(a) A linearly polarized pump beam is
shone on a spatial light modulator (SLM) to generate the pump lattice pattern.
The emission is filtered and projected on a monochromator in tandem with a
streak camera for time-resolved measurements. (b) Michelson interferometer with a
retroreflector in one arm is used to interfere the emission from one condensate
with the other. The interference is sent to a streak camera for the time resolution.}
\label{fig:setup}
\end{figure}

The details of the microcavity structure could be found in
Ref.~\citenum{kammann_crossover_2012} and the optical setup is shown in detail
in Figure~\ref{fig:setup}. The sample was cooled to
$\sim 10\ \mathrm{K}$ using a cold-finger cryostat. The wedged structure of the
microcavity allows accurate probing of the cavity and quantum-well detuning
($\sim -9\ \mathrm{meV}$). The sample was excited nonresonantly into the Bragg
mode ($\sim 0.1\ \mathrm{eV}$ above the cavity mode) by 180 fs pulses to make
sure that the phase of the excitation laser was lost by multiple relaxations
towards the ground state. The condensate lattice was created using a reflective
spatial light modulator (SLM)~(see Fig.~\ref{fig:setup}b). A high numerical aperture
microscope objective (NA = 0.7) focuses the laser ($\lambda \simeq
730~\mathrm{nm}$) to a $\sim 1.3\ \mathrm{\mu m}$ diameter spot and collects the
PL. Imaging the Fourier plane of the emission to a spectrometer allows mapping
out of the energy-momentum space.

\subsection{Gross-Pitaevskii equation with Langevin noise}
\label{sec:GP}

To simulate the stochastic behaviour of our coupled condensates, we use the
mean-field Gross-Pitaevskii equation~\cite{wouters_spatial_2008} given by
\begin{equation}
  \label{eq:gp}
  i\hbar \frac{\partial \psi(\mathbf r)}{\partial t}=\Big\{ E_0 -\frac{\hbar^2}{2m}
  \nabla^2_\mathbf{r}+\frac{i \hbar}{2}
  [R[n_R(\mathbf{r})]-\gamma_c]+\hbar g
  \vert\psi(\mathbf{r})\vert^2+V_R(\mathbf r)
\Big\} \psi(\mathbf r)+f(t),
\end{equation}
where $E_0=\hbar \omega$, $m$ is the effective mass, $R$ is the incoming rate of
polaritons from a hot exciton reservoir with a local density of $n_R$ to the condensate,
$\gamma_c$ is the decay rate of polaritons, $g$ is the repulsion constant
accounting for polariton-polariton interactions, and $V_R$ is the repulsive
potential created by the nonresonant excitation pump. The mean-field repulsive potential $V_R$ is
given by the linear expression
\begin{equation}
  V_\mathbf{r} = \hbar g_R n_R(\mathbf r)+\hbar \mathcal{G} P(\mathbf r),
\end{equation}
where $P$ is the spatially dependent pumping rate of excitons in the reservoir,
and $g_R$ and $\mathcal{G}$ are phenomenological constants given by the
experimental values. Similar to Ref.~\citenum{read_stochastic_2009}, $f(t)$ is the
Langevin noise given by the correlator
\begin{equation}
  \langle f(t) f^*(t') \rangle = \frac{1}{2} R n_R(\mathbf r) \delta(t-t').
\end{equation}
Equation~\ref{eq:gp} is coupled to a rate equation for the reservoir given by:
\begin{equation}
  \dot{n}_R(\mathbf r)=P(\mathbf r)-\gamma_Rn_R(\mathbf r)-Rn_R(\mathbf r)\vert
  \psi \vert^2.
\end{equation}
Here, $\gamma_R \gg \gamma_c$ is the decay rate of the excitons, and the last
term accounts for depletion of the reservoir due to stimulated emission
into the condensate. It is worth mentioning that we do not assume any energy
relaxations in any of our simulations.

We use a fifth-order Adams-Bashforth-Moulton predictor-corrector method and take
an average for 75 realizations using these parameters: $\hbar
R_r=\unit[0.05]{\mu m^{-2} meV}$, $\hbar g_R=0$, $\mathcal{G}=\unit[0.0175]{\mu
m^2}$, $\hbar g=\unit[0.02]{meV \mu m^2}$, $\gamma_c=\unit[0.5]{ps^{-1}}$,
$\gamma_R=\unit[0.01]{ps^{-1}}$. $P$ is set to give the observed blueshift in
the experiment.

\subsection{The dissipative coupling mechanism}
\label{si:coupling}
\begin{figure}
  \centering
  \includegraphics[height=0.5\textwidth]{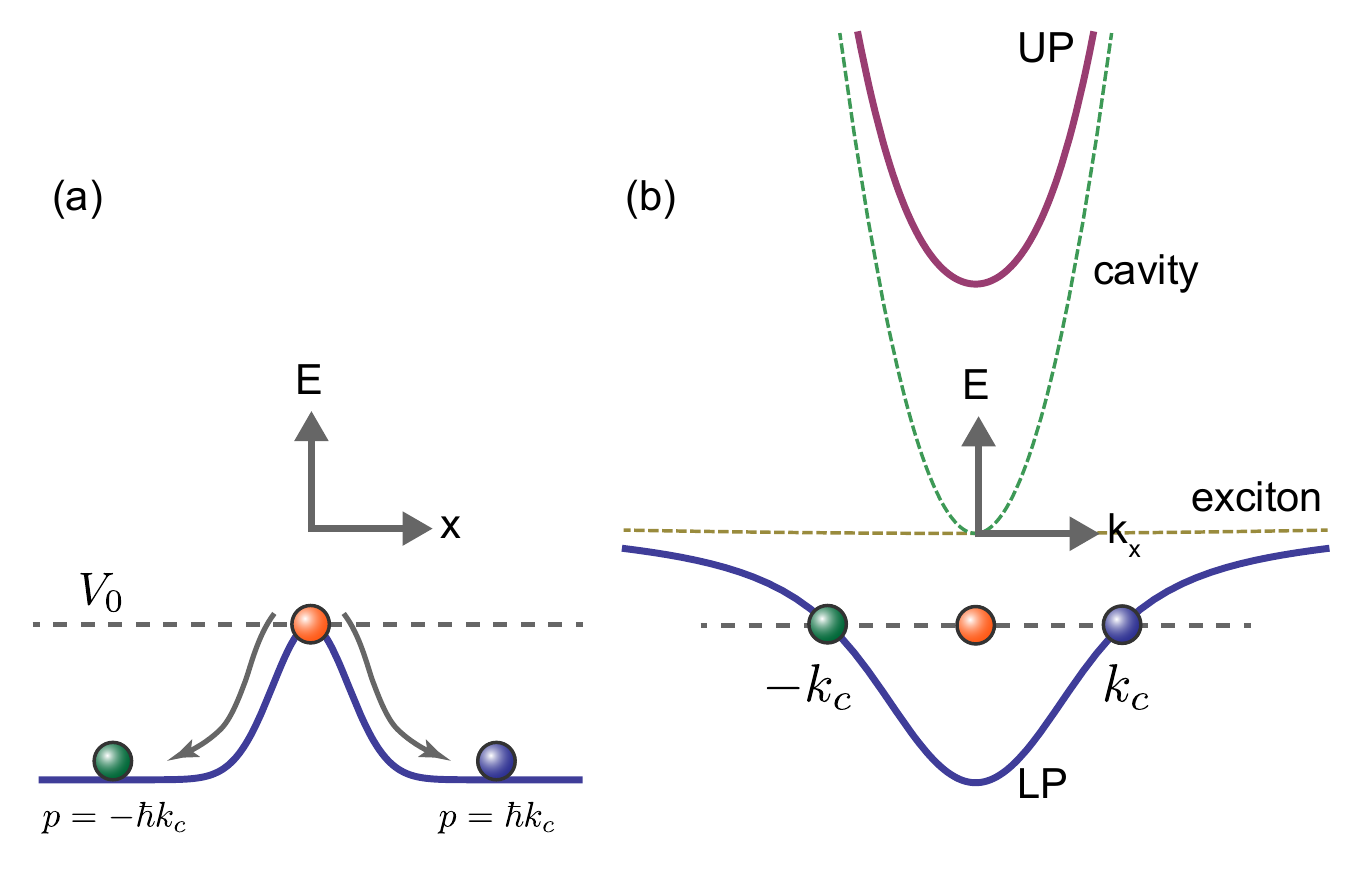}
  \caption{ Due to the repulsive
interaction of polaritons with background reservoir, the condensate is
formed on top of a potential $V_0$. Polaritons roll down the potential and gain an
in-plane momentum $p=\hbar k_c=\sqrt{2mV_0}$.}
  \label{fig:blueshift-schem}
\end{figure}

To gain more quantitative insight into the dependence of losses on the distance between the 
pair of condensates we consider the tail $\tilde{\Psi}(\mathbf{r})$ of the condensate wave 
function in the region away from the excitation spots.
In this intercondensate region the polaritons move as free particles and $\tilde{\Psi}(\mathbf{r})$
can be written as superposition of tails from individual condensates. 
For two condensates of equal size separated by a distance $\mathbf{a}$ we have
\begin{equation}
 \label{eq:decay_rate_bessel}
 \tilde{\Psi}(\mathbf{r})=\frac{1}{\sqrt{2}}\left[
 \tilde{\psi}\left(\mathbf{r}+\frac{\mathbf{a}}{2}\right)
 +e^{i\theta}\tilde{\psi}\left(\mathbf{r}-\frac{\mathbf{a}}{2}\right)\right] ,
\end{equation}
where $\theta $ is the phase difference between the two condensates.
The total number of emitted polaritons during the condensate formation is given by
\begin{equation}
\label{eq:lifetime-integral}
 I  =\int\frac{\mathrm{d}^2k}{(2\pi)^2}
  |\tilde{\Psi}(\mathbf{k})|^{2},
\end{equation}
\begin{equation}
  \tilde{\Psi} (\mathbf{k}) =\int \mathrm{d}^{2}re^{-i\mathbf{k}\cdot \mathbf{r}}\tilde{\Psi}(\mathbf{r})
  =\frac{\tilde{\psi}(k)}{\sqrt{2}}\left[
  e^{i\mathbf{k}\cdot \frac{\mathbf{a}}{2}}+e^{i\theta }e^{-i\mathbf{k}\cdot \frac{\mathbf{a}}{2}}
                          \right].
\end{equation}
Due to the interactions between polaritons and the exciton reservoir, the condensate is
blueshifted in energy. Outside the pump spot, this potential energy is converted
to kinetic energy with a specific in-plane wavevector $k_c$. As a result, the
tail of each condensate is represented in the reciprocal space by a ring with
the wave vector $k_c$ satisfying $\omega_\mathrm{LP}(k_c)=\omega_c$, where
$\omega_\mathrm{LP}(k)=\hbar k^2/2m^*$ is the dispersion of the microcavity mode and
$\omega_c$ is the energy of the
condensate~\cite{wouters_spatial_2008,kammann_nonlinear_2012} (see also
Supplementary Fig.~\ref{fig:blueshift-schem}). Since the condensate wavefunction
for outflowing polaritons is mostly composed of those at $k=k_c$, it can be
approximated by $\vert \tilde{\psi}(k) \vert^2 \propto \delta (k-k_c)$.  Substituting
this in Eq.~\ref{eq:lifetime-integral} gives the number of emitted polaritons
\begin{equation}
 \label{eq:tau_of_theta}
 I(k_ca,\theta) \propto [1 +J_{0}(k_{c}a) \cos(\theta)] ,
\end{equation}
where $J_0$ is the Bessel function.

It is easy to show that in the general case of $N$ condensates placed at
positions $\mathbf{R}_n$ and having phases $\theta_n$ the total loss function
is given by
\begin{equation} I\propto \left[ 
        1+\frac{2}{N}\sum_{n,n'} \cos(\theta_n-\theta_{n'})J_0(k_c
        \vert \mathbf{R}_n-\mathbf{R}_{n'}\vert)
        \right], 
\end{equation} 
where the summation is over all $N(N-1)/2$ pairs. In the case of an equilateral
triangular lattice the loss function is minimized when either all three neighbours are in
phase or when there is a $2 \pi/3$ phase difference between the neighbours. The
nontrivial case where there is a $2\pi/3$ phase difference going clockwise or
anticlockwise between the neighbours corresponds to two topologically different
vortex states with the winding number of +1 or -1.

\begin{figure}
\centering
\includegraphics{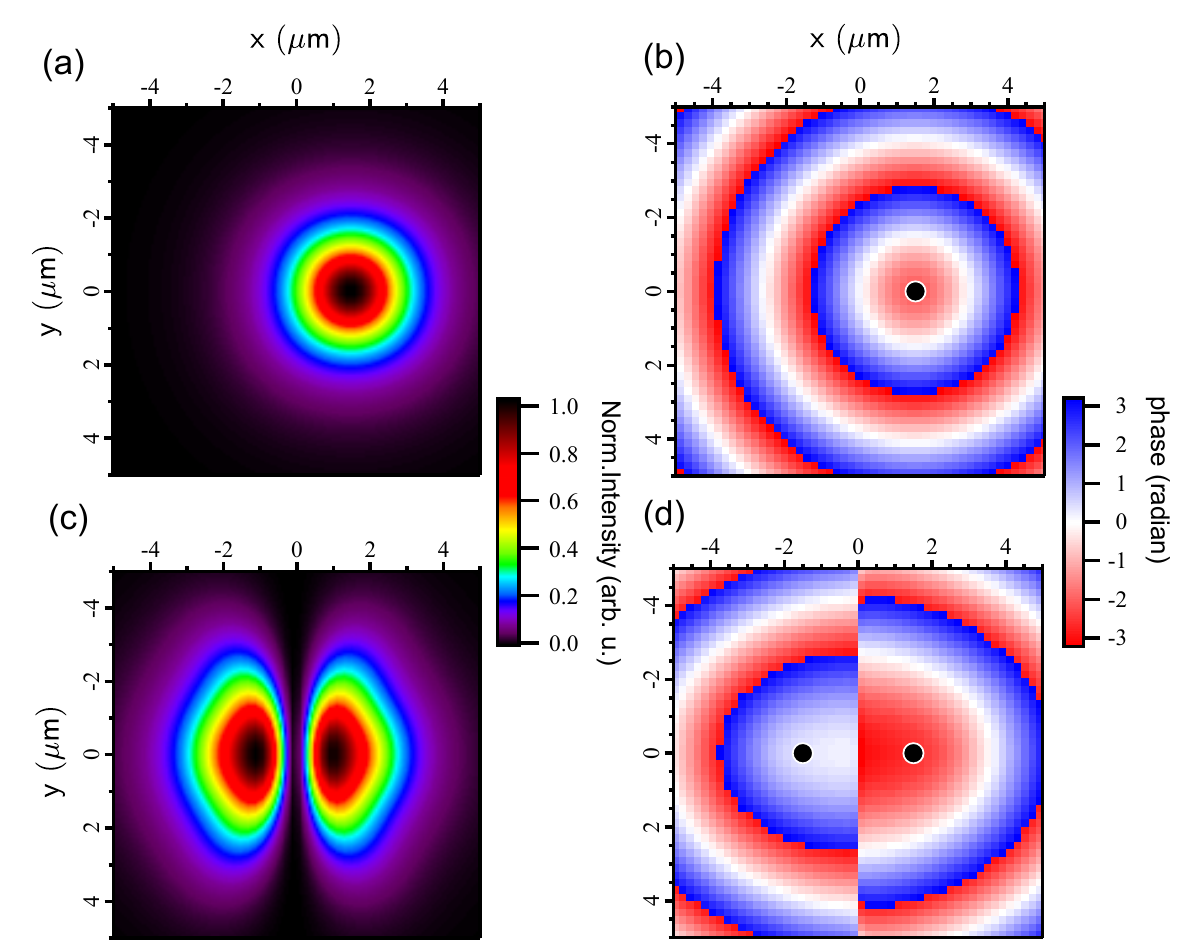}
\caption{
    (a,b) The intensity and phase diagram of a single condensate is shown at
  $x=\unit[1.5]{\mu m}$. The flowing polaritons at
  $x=\unit[-1.5]{\mu m}$ are in-phase with the center of the condensate.
  (c,d) The intensity and phase diagram of a phase-locked
  condensate is shown with a separation distance of $\unit[3]{\mu m}$. The coupled
  condensates have opposite phases at the centers, opposite to the phase of the flowing
  polaritons from the other condensate.
}
\label{fig:phase-diagram}
\end{figure}

It is important to note the difference between the dissipative coupling
mechanism described here and the ``coherent ballistic coupling'' mechanism described
in the previous publications~\cite{tosi_geometrically_2012}. The dissipative
coupling mechanism causes two condensates with a separation smaller than
$L=\pi/k_c$ to couple anti-phase (see Figure~\ref{fig:phase-diagram}), whereas
the ballistic coupling would always provide for the in-phase coupling.

\begin{figure}[p]
\centering
\includegraphics[width=1\textwidth]{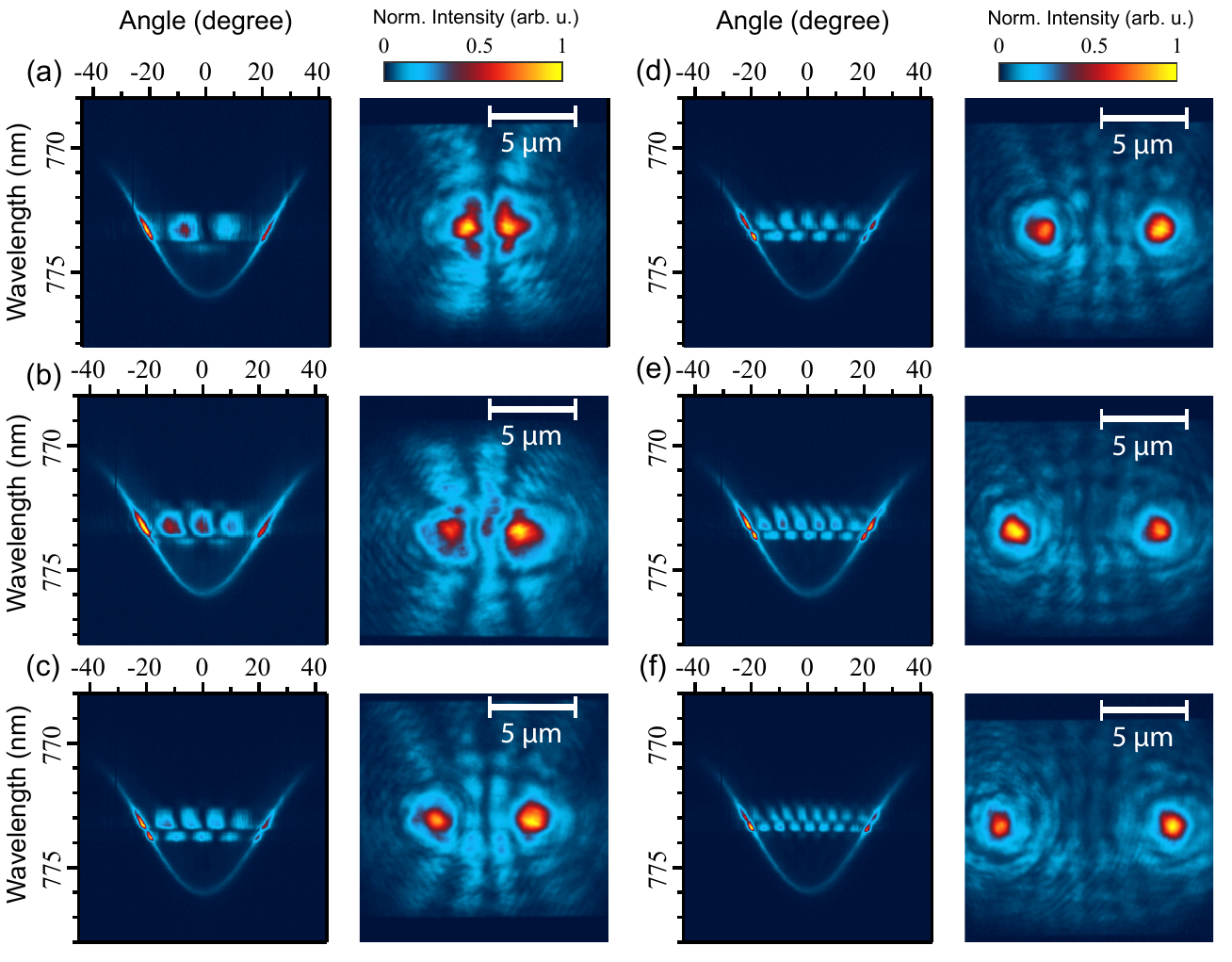}
\caption{The real space and the
momentum space of the two condensates for the separation distance of
$\unit[2.2]{\mu m}$ (a), $\unit[4]{\mu m}$ (b), $\unit[5.5]{\mu m}$ (c),
$\unit[6.5]{\mu m}$ (d), $\unit[8.2]{\mu m}$ (e) and $\unit[9.7]{\mu m}$ (f) are
shown.}
\label{fig:ks-rs}
\end{figure}

\begin{figure}[p]
\centering
\includegraphics[height=1\textwidth]{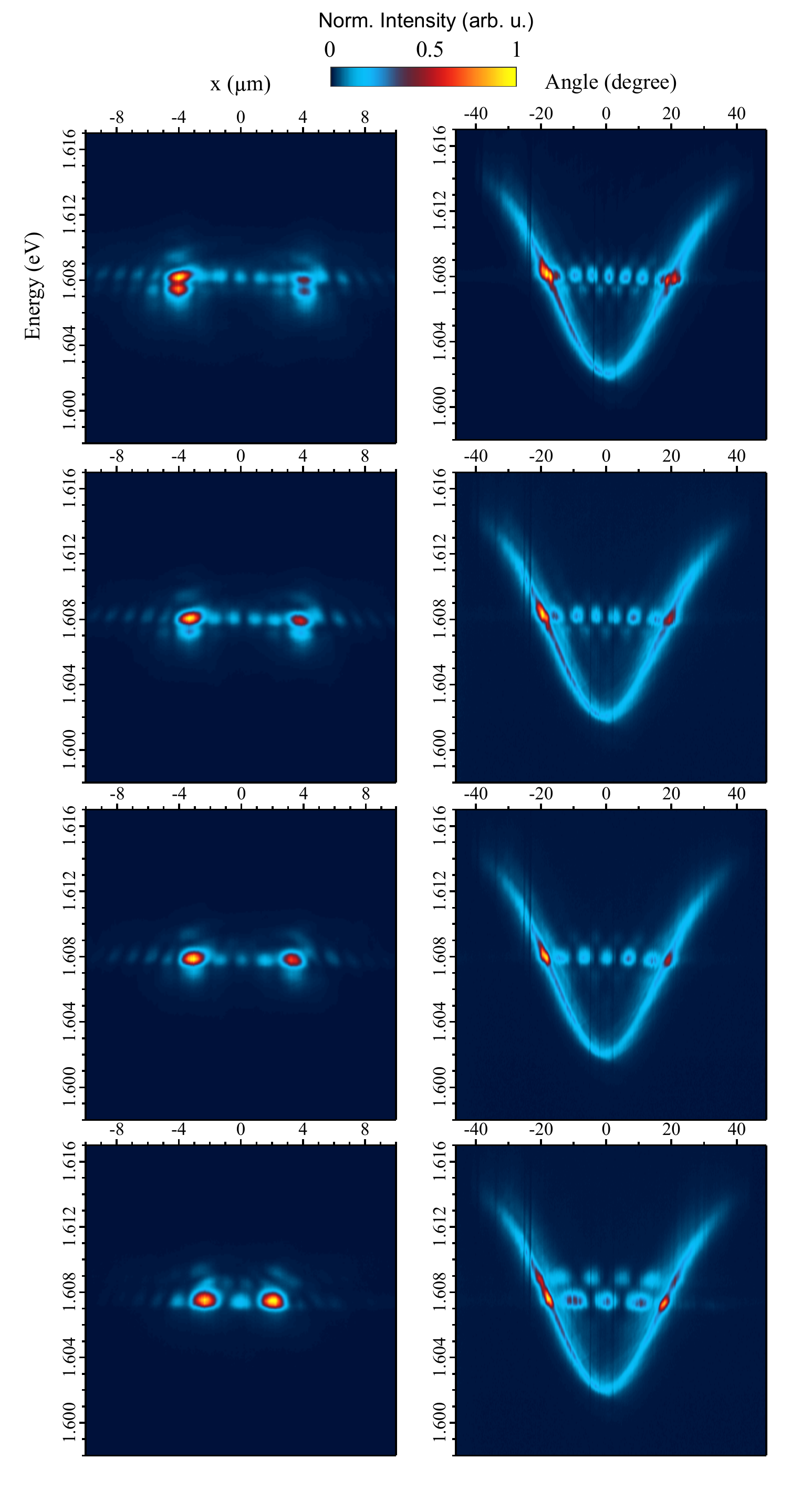}
\caption{The energy-resolved real space and the
momentum space images of the two condensates for different separations are
shown. The real spaces images show that the fringes in between the condensates
are at the same energy as the condensates are.}
\label{fig:rs-er}
\end{figure}

\begin{figure}
\centering
\includegraphics[width=1\textwidth]{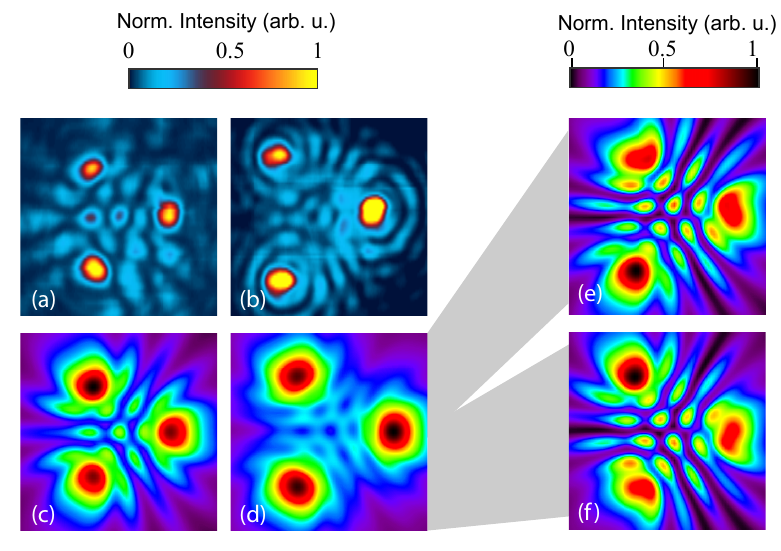}
\caption{
    (a,b) The intensity pattern of an all-in-phase
  triangular lattice, and a vortex state is shown respectively. Th GP
  simulations for 75 realizations are shown in (c) and (d). Simulations confirm
  that the vortex states are composed of clockwise and anticlockwise phase
  vortices shown in (e) and (f).
}
\label{fig:3beams-larger-sep}
\end{figure}

\begin{figure}
\centering
\includegraphics[width=1\textwidth]{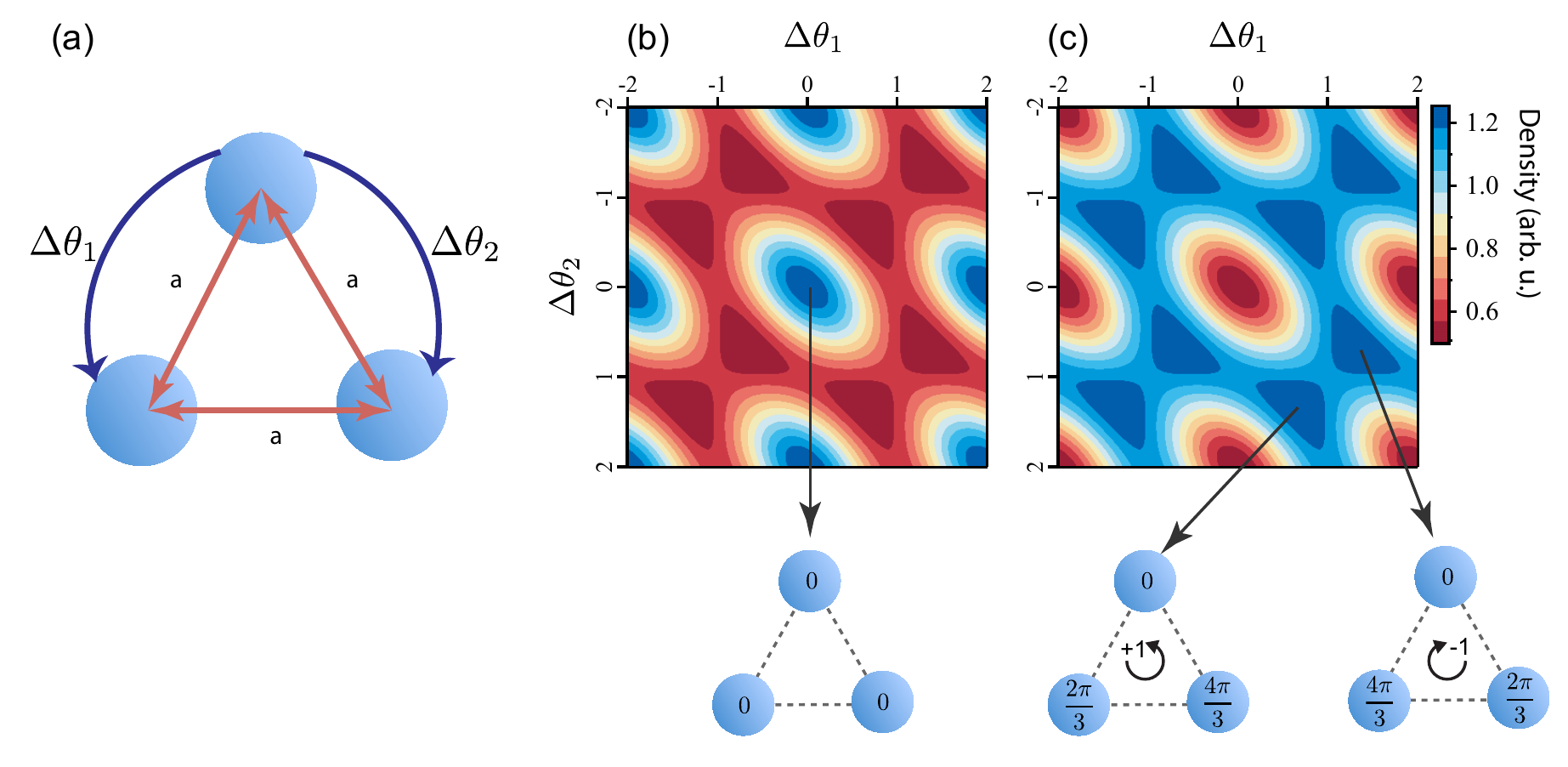}
\caption{
    (a) An equilateral triangular lattice with a separation of $a$
  between the neighbours and phase difference of $\Delta \theta_1$ and
  $\Delta \theta_2$ between the neighbours is shown. (b,c) The
  density function for different phases between the condensates is shown for
  $k_c=\unit[1]{(\mu m)^{-1}}$ and $a=\unit[7]{\mu m}$ (b) and $a=\unit[10]{\mu
  m}$ (c). The condensates fall into a state with the highest density at the
  onset of condensation. The maximum density in (c) is composed of two
  topologically different states, which correspond to clockwise or anticlockwise
  $2 \pi/3$ phase difference between the sites.
}
\label{fig:3beams-geometry}
\end{figure}

\begin{figure}
\centering
\includegraphics[width=0.75\textwidth]{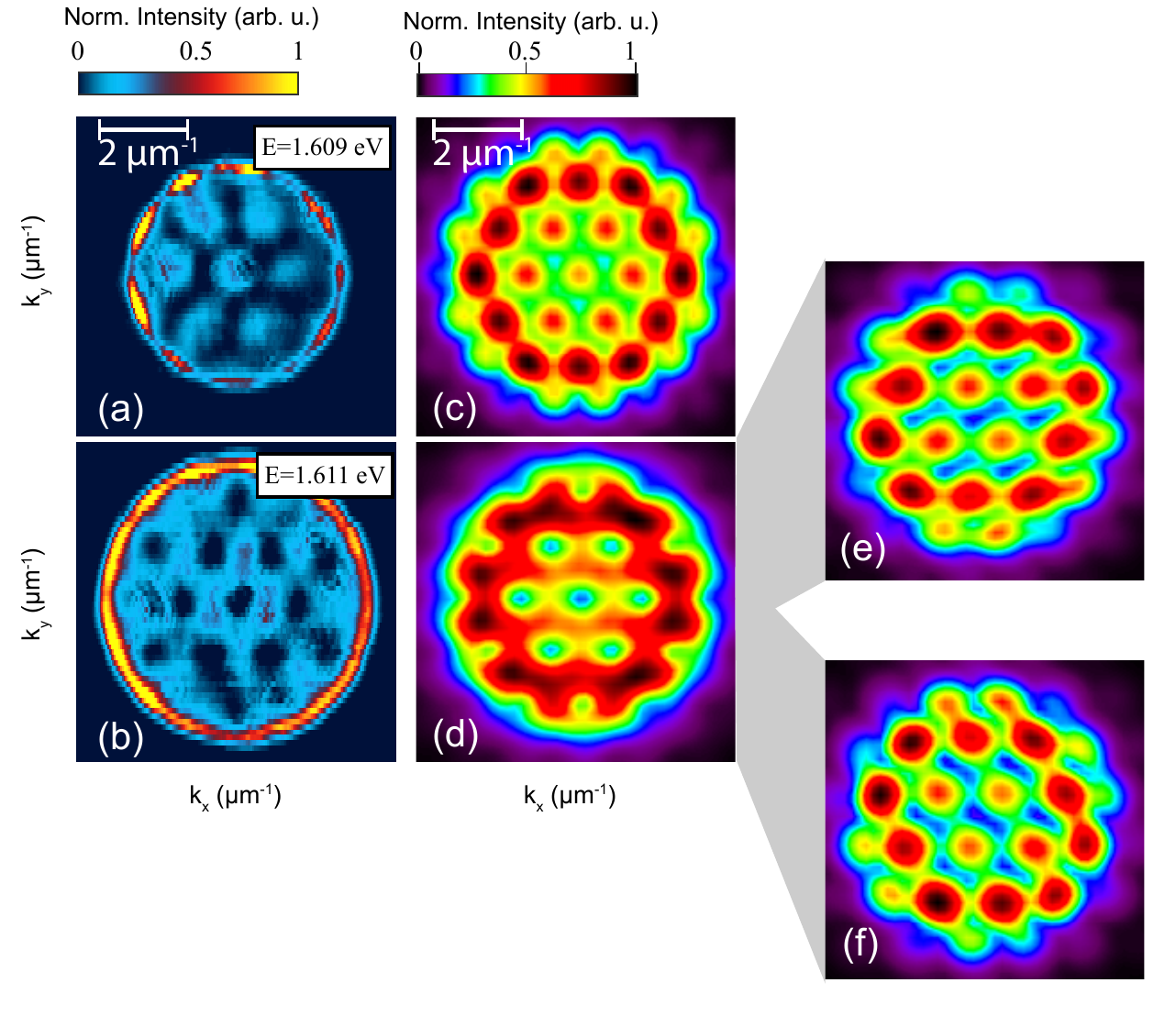}
\caption{(a,b) The momentum space of
a triangular condensate lattice for a lattice constant of $\unit[5.5]{\mu m}$ is
shown at two different energies, and wavevectors. (a) corresponds to the case where all
three condensates are in phase and (b) shows the case where they are in a
mixture of clockwise and anticlockwise vortex states with winding numbers of
$\pm1$. (c,d) The GP simulations with Langevin noise with 75 realizations are shown for a separation
of $\unit[6.5]{\mu m}$ (c) and $\unit[5.5]{\mu
    m}$ (d). (e,f) The simulations show that the honeycomb
pattern observed in (d) is indeed a mixture of clockwise (e) and anticlockwise
(f) vortex states where the phase difference between neighbors is $2 \pi/3$.}
\label{fig:3cond-kspace}
\end{figure}

\clearpage

\end{document}